\documentstyle[12pt]{article}
\newcommand{\sect}[1]{\setcounter{equation}{0}\section{#1}\indent}
\renewcommand{\theequation}{\thesection.\arabic{equation}}
\renewcommand{\thefootnote}{\fnsymbol{footnote}}
\newcommand{\EQ}{\begin{equation}}
\newcommand{\EN}{\end{equation}}
\newcommand{\bea}{\begin{eqnarray}}
\newcommand{\ena}{\end{eqnarray}}
\newcommand{\vs}[1]{\vspace{#1 mm}}

\renewcommand{\a}{\alpha}
\renewcommand{\b}{\beta}
\renewcommand{\c}{\gamma}
\renewcommand{\d}{\delta}

\newcommand{\G}{\Gamma}

\newcommand{\pa}{\partial}

\newcommand{\uda}{\nearrow \kern-1em \searrow}

\newcommand{\la}{\lambda}
\newcommand{\La}{\Lambda}

\makeatletter
\def\eqnarray{%
 \stepcounter{equation}%
 \let\@currentlabel=\theequation
 \global\@eqnswtrue
 \global\@eqcnt\z@
 \tabskip\@centering
 \let\\=\@eqncr
 $$\halign to \displaywidth\bgroup\@eqnsel\hskip\@centering
 $\displaystyle\tabskip\z@{##}$&\global\@eqcnt\@ne
 \hfil$\displaystyle{{}##{}}$\hfil
 &\global\@eqcnt\tw@$\displaystyle\tabskip\z@{##}$\hfil
 \tabskip\@centering&\llap{##}\tabskip\z@\cr}
\makeatother

\begin{document}
\begin{titlepage}
\setcounter{page}{0}
\begin{flushright}
EPHOU 97-005\\
May 1997
\\
\end{flushright}

\begin{center}
{\Large Seiberg-Witten Theory of Rank Two Gauge Groups\\
 and
Hypergeometric Series}

\vs{3}
{\large
Takahiro
Masuda
\footnote{e-mail address: masuda@phys.hokudai.ac.jp},\\
Toru Sasaki
\footnote{e-mail address: sasaki@phys.hokudai.ac.jp}
\\
and \\
Hisao Suzuki\footnote{e-mail address: hsuzuki@phys.hokudai.ac.jp}}\\
\vs{3}
{\em Department of Physics, \\
Hokkaido
University \\  Sapporo, Hokkaido 060 Japan} \\
\end{center}
\vs{2}

\centerline{{\bf{Abstract}}}
In $SU(2)$ Seiberg-Witten theory, it is known that the dual pair of 
fields are expressed by hypergeometric functions. 
As for the theory with $SU(3)$ gauge symmetry without 
matters, it was shown that the 
dual pairs of fields can be expressed by means of the Appell function of type 
$F_4$. These expressions are convenient for analyzing analytic properties 
of fields. We investigate the relation between Seiberg-Witten theory of 
rank two gauge group without matters and hypergeometric series of 
two variables. 
It is shown that the relation between gauge theories and Appell 
functions can be observed for other classical 
gauge groups of rank two.  For $B_2$ and 
$C_2$, the fields are written in terms of Appell functions 
of type $H_5$. For $D_2$, we can express fields by Appell functions of type 
$F_4$ which can be decomposed 
to two hypergeometric functions, corresponding to the 
fact $SO(4) \sim SU(2)\times SU(2)$. We also consider the integrable curve 
of type $C_2$ and show how the fields are expressed by Appell 
functions. However in the case of exceptional group $G_2$, our 
examination shows that 
they can be  represented by hypergeometric series which 
does not correspond to the Appell functions. 

\end{titlepage}
\newpage

\renewcommand{\thefootnote}{\arabic{footnote}}
\setcounter{footnote}{0}

\sect{Introduction} 

Many recent researches which have been originated to the work of 
Seiberg-Witten \cite{SW}, 
are now clarifying the non-perturbative aspect of various 
supersymmetric gauge field theories through the 
duality symmetry. Seiberg and Witten gave the exact solution of 
the low energy effective theory with gauge group $SU(2)$ without 
matters, which provide a kind of understanding of the confinement 
through the monopole condensation. Moreover it was pointed out that there 
exist some special points in the moduli space of the extended theories 
constructed by introducing the matter fields and taking higher rank 
gauge symmetry, 
which realize the superconformal field theories \cite{AD,APSW,EHIY}. 
Such possibilities of generalization have been studied extensively 
and the frameworks of the extended theories have 
been  constructed elegantly 
\cite{KLTY,APS,HO,DS,Ha,AS,CDF,KLerT,BSDKM}. 

The integrability of such known Seiberg-Witten type theories 
are insured by the relation to the integrable system \cite{NTMIM}, which
shows that the hyperelliptic curve of the Seiberg-Witten model 
corresponds to the spectral curve of the periodic Toda lattice 
\cite{MW}. It is 
possible to construct the Seiberg-Witten model with the 
classical gauge group including 
 some cases of the exceptional gauge group \cite
{AAMLPG,AAG,LW}, and also the model 
which has no corresponding gauge group, all by using 
 suitable spectral curve of integrable systems \cite{MW}. 
Recently it was pointed out that 
 equivalent curves are obtained by the 
 fibrations of ALE spaces of type I$\!$I 
superstring theories conpactified on Calabi-Yau threefolds 
\cite{KKLMV,KLMVW,Ler,Kle}, although such equivalence is not manifest 
but physical in the case of the exceptional group \cite{LW,Bro}. 

As for the explicit evaluation of the theory, 
 various methods to obtain the physical information in 
such Seiberg-Witten models have been investigated. 
 One major way is to use the concrete expression of 
 dual Higgs pairs obtained 
by solving the Picard-Fuchs equation \cite{IMNSA} or 
 evaluating the integral representation \cite{DKP,MS2}, to 
get the low energy effective action. 
 There are also direct calculations 
for the prepotential \cite{Ma} which have been developed in relating 
the soliton theory \cite{EYIMBM} in the weak coupling region. 
 These result have been almost consistently checked 
by the direct instanton calculus in the weak coupling region \cite{Inst}, 
 especially have shown  good coincidence in the case 
without matters in the classical gauge group \cite{IS}.
The evaluations for the prepotential 
 have been carried out even in the strong coupling region \cite{DP}. 
 In the case of $SU(2)$, besides the prepotential, the BPS spectrum 
have been analyzed in both regions by means of global quantum symmetry
 \cite{BF}. Soft breaking of $N=2$ supersymmetry with massless or  
 massive hypermultiplets has been investigated 
 due to the strong coupling analysis \cite{AGZM}.  

Although physical information can be given by 
the prepotential, in this article we concentrate on obtaining analytic 
 expressions of 
dual Higgs pairs exactly to 
investigate their analytic properties. 
 It is known that 
they should be represented by hypergeometric series in the classical 
gauge group \cite{DKP,MS2}, and obtained order 
by order recursively by analyzing the Picard-Fuchs equations  
\cite{KLerT,IY,EFT,Oh,IS}. 
We choose an alternative approach to evaluate the integral representation 
 which has been investigated among our previous works \cite
{MS2,MS}. In the case of $SU(n)$ and $SO(2n)$ theory, results obtained by 
this method are expressed by using hypergeometric series \cite{MS2}, 
however this representation does not seem to 
be convenient to see analytic behaviors for 
generic region of the branch. 
In other words, we have to perform the expansion in various regions 
where we can not obtain the result expanded 
in terms of the hypergeometric series. 
However in the case of $SU(3)$ \cite{KLerT}, 
the analytic property is shown to be 
apparent because the solution can be written 
by Appell functions. These functions are second order 
hypergeometric series in two variables, and can be represented 
by using the simple hypergeometric functions 
in terms of each variable, so that 
we can obtain the expression close to the boundary to the 
convergence region. 
 Therefore the rank two gauge groups are expected to be special cases where 
solutions can be represented by analogous hypergeometric functions 
in a uniform way so that their analytic properties 
can be seen manifestly.

Our aim in this article is to clarify the relation 
between hypergeometric series and $N=2$ supersymmetric Yang-Mills theories 
with rank two gauge group both by use of explicit evaluations and 
by the Picard-Fuchs equations.  
As the result, we will show that the dual pairs of fields are represented 
 by Appell functions for the classical gauge groups of rank two. 
In the case of exceptional group $G_2$, the 
solution of the dual pairs of fields have been obtained 
not in the form of hypergeometric series in \cite{I}. 
 In this article we will find that we can obtain the 
expression in the form of hypergeometric series. 
However they turn out to be the third order hypergeometric series, which means 
that they are not within the Appell system in this case. Above 
solutions obtained by explicit evaluations can satisfy 
the Picard-Fuchs equations of these 
theories as we will see in this article. 

This article is organized as follows. In section two, we will review the 
theory with gauge group $SU(3)$ and will show how to obtain 
 the result by Klemm {\it et al.}\rm \cite{KLerT}
by explicit evaluations of the integral. In section three, we will calculate 
in the case of $SO(5)$ and $Sp(4)$. In section four we will deal with the
theory with $SO(4)$, so that we see that results reflect 
the fact $SO(4)\sim SU(2)\times SU(2)$. 
 In section five we will consider 
the integrable system $C_2$ whose counterpart gauge 
group is not known now. From sect. 3 to sect. 5, 
the results can be represented by 
suitable Appell functions with various parameterizations. 
In section six, we will deal with exceptional group $G_2$ to show 
that the solution can be represented by the generalized hypergeometric 
function of order 3 in terms of a special parameterization of 
the moduli parameter. Last section will be  devoted some discussion. 


\sect{Review of $SU(3)$ gauge theory}

To begin with, we review $SU(3)$ gauge theory
whose dual pairs were previously evaluated exactly by Klemm, Lerche and 
Theisen  \cite{KLerT} 
by means of solving the Picard-Fuchs equation directly.
 We are going to evaluate dual pairs of  $SU(3)$ gauge theory 
by integrating meromorphic differential
$\lambda$ explicitly in the weak coupling region
to see that we can recover the exact solutions. 
 Hyperelliptic curve and meromorphic differential
$\lambda$ of $SU(3)$ gauge theory 
are given by
\bea
y^2=W^2-\Lambda^6,\    W=x^3-ux-v, 
\ena
\bea
\lambda=\frac{xW^\prime}{y}dx. 
\ena
Higgs field $a_i$ and its dual $a_D^i\ (i=1,2,3)$ are given by
\bea
a_i=\oint_{\a_i}\la,\  a_D^i=\oint_{\b_i}\la, 
\ena
where $\sum_{i=1}^3a_i=0$, 
and $\a_i,\b_i \ (i=1,2,3)$ are basis of homology cycle. In
 the weak coupling region $(\La \sim 0)$,
we expand $1/y$ of $\la$ with small $\La$ \cite{MS2,MS}
\bea
\la
&=&dx\int^{+i \infty}_{-i\infty}
\frac{ds}{2\pi i}\frac{\Gamma(-s)\Gamma(\frac{1}{2}+s)}
{\Gamma(\frac{1}{2})\ 2s}
(-\Lambda^6)^s W^{-2s},
\ena
where we introduce $s$ integral which 
picks up poles at $s=0,1,2,\cdots$, in place of summation, 
and integrate by parts. The detailed calculation is given in 
 Appendix A. We first perform the contour integral along 
$\alpha_i$ and $\beta_i$ cycle, 
next evaluate s integral. 
In the weak coupling region the integral of $\la$ 
along $\alpha_i$ cycle picks up simple poles of (2.4). 
In the region $u\sim \infty$, 
 we take $\alpha_1(\alpha_2)$ cycle to enclose two point 
$x\sim \sqrt{u}(-\sqrt{u})$ and $x\sim v/u
\sim 0$, so as to be same basis taken by Klemm \it et al.\rm \cite{KLerT}. 
 Moreover this contour integration can 
be carried out by line integration from $x=0$ to $x=\sqrt{u}(-\sqrt{u})$ 
 supplemented by $\sin 2s\pi/\pi$. Similarly the integration 
along $\beta_i$ cycle can be replaced to this line integration 
without multiplied 
by $\sin 2\pi s$ because this contour initially 
intersects the blanch cut. Since the reduction of 
two different roots of the curve (2.1) 
to one classical root caused by $\La\rightarrow 0$,  
provides an excessive contribution 
 $\alpha_i/2$ to $\beta_i$, we have to subtract this. Therefore \cite{MS2,MS}
\bea
a_1&=&\int_0^{\sqrt u}\la{\sin 2\pi s\over \pi},\ \ \  
a_D^1=\frac{1}{2\pi i }\int_0^{\sqrt u}\la-{1\over 2}a_1 ,\\
a_2&=&\int_0^{-\sqrt u}\la{\sin 2\pi s\over \pi} ,\ \ \ 
a_D^2=\frac{1}{2\pi i }\int_0^{-\sqrt u}\la -{1\over 2}a_2,
\ena
To be concrete, we expand $W^{-2s}$ by $1/(x^3-ux)$, and 
integrate  $\la$ by $x$ from $x=0$ to $x=\sqrt u \  (x=-\sqrt u)$ and
 evaluate $s$ integration. 
After some arrangement we obtain    
\bea
a_1&=&\sqrt u F_4\bigg(-\frac{1}{6},\frac{1}{6},1,\frac{1}{2};
\frac{27\La^6}{4u^3},\frac{27v^2}{4u^3}\bigg)
+\frac{v}{2u} F_4 \bigg(\frac{1}{3},\frac{2}{3},1,\frac{3}{2};
\frac{27\La^6}{4u^3},\frac{27v^2}{4u^3}\bigg)
,\label{eq:A2a1}\\
a_D^1&=&\frac{1}{2\pi i}
\sqrt u \sum_{m,n}\frac{(-\frac{1}{6})_{m+n}
(\frac{1}{6})_{m+n}}
{m!n!(1)_m(\frac{1}{2})_n}
\bigg(\frac{27\La^6}{4u^3}\bigg)^m
\bigg(\frac{27v^2}{4u^3}\bigg)^n \nonumber \\
& &\ \ \times \bigg[
-2\psi(m+1)+\psi(-\frac{1}{6}+m+n)+\psi(\frac{1}{6}+m+n)
+\log\bigg(-\frac{27\La^6}{4u^3}\bigg)\bigg] \nonumber \\
& &+\frac{1}{2\pi i}
\frac{v}{2u} \sum_{m,n}\frac{(\frac{1}{3})_{m+n}
(\frac{2}{3})_{m+n}}
{m!n!(1)_m(\frac{3}{2})_n}
\bigg(\frac{27\La^6}{4u^3}\bigg)^m
\bigg(\frac{27v^2}{4u^3}\bigg)^n \label{eq:A2aD1}\\
& &\ \ \times \bigg[
-2\psi(m+1)+\psi(\frac{1}{3}+m+n)+\psi(\frac{2}{3}+m+n)
+\log\bigg(-\frac{27\La^6}{4u^3}\bigg)\bigg],\nonumber 
\ena
\bea
a_2&=&-\sqrt u F_4\bigg(-\frac{1}{6},\frac{1}{6},1,\frac{1}{2};
\frac{27\La^6}{4u^3},\frac{27v^2}{4u^3}\bigg)
+\frac{v}{2u} F_4 \bigg(\frac{1}{3},\frac{2}{3},1,\frac{3}{2};
\frac{27\La^6}{4u^3},\frac{27v^2}{4u^3}\bigg)
,\\
a_D^2&=&-\frac{1}{2\pi i}
\sqrt u \sum_{m,n}\frac{(-\frac{1}{6})_{m+n}
(\frac{1}{6})_{m+n}}
{m!n!(1)_m(\frac{1}{2})_n}
\bigg(\frac{27\La^6}{4u^3}\bigg)^m
\bigg(\frac{27v^2}{4u^3}\bigg)^n \nonumber \\
& &\ \ \times \bigg[
-2\psi(m+1)+\psi(-\frac{1}{6}+m+n)+\psi(\frac{1}{6}+m+n)
+\log\bigg(-\frac{27\La^6}{4u^3}\bigg)\bigg] \nonumber \\
& &+\frac{1}{2\pi i}
\frac{v}{2u} \sum_{m,n}\frac{(\frac{1}{3})_{m+n}
(\frac{2}{3})_{m+n}}
{m!n!(1)_m(\frac{3}{2})_n}
\bigg(\frac{27\La^6}{4u^3}\bigg)^m
\bigg(\frac{27v^2}{4u^3}\bigg)^n \\
& &\ \ \times \bigg[
-2\psi(m+1)+\psi(\frac{1}{3}+m+n)+\psi(\frac{2}{3}+m+n)
+\log\bigg(-\frac{27\La^6}{4u^3}\bigg)\bigg],\nonumber 
\ena
where $(\a)_m=\G(a+m)/\G(a),\psi(x)=\G^\prime(x)/\G(x)$
and defining expression of $F_4$ function is given by
\bea
F_4(\a,\b,\c,\d;x,y)
=\sum_{m,n}\frac{(\a)_{m+n}(\b)_{m+n}}{m!n!(\c)_m(\d)_n}x^my^n.
\ena
This is the Appell $F_4$ function 
which are order two hypergeometric series in terms of each variables 
\cite{HTF}. The solutions (2.7)-(2.10) 
are equal to the results obtained by Klemm \it et al.\rm 
 \cite{KLerT} up to over all sign of $a_2,a_D^2$.

Analytic property of $F_4$ function is apparent 
because this is a natural extension of Gaussian sum 
in two variables and is expressed by using 
 hypergeometric functions of $x$ as
\bea
F_4(\a,\b,\c,\d;x,y)
=\sum_{n}\frac{(\a)_{n}(\b)_{n}y^n}{(\d)_n\,n!}F(\a+n,\b+n,
\c,x),
\ena
where the hypergeometric function $F(a,b,c;z)$ is defined by
\bea
F(a,b,c;z)=\sum_{n}{(a)_n(b)_n\over (c)_n\,n!}\ z^n,
\ena
and similar manipulation can be carried out in terms of $y$.  
Therefore it is easy to know the behavior near the boundary of the 
convergence region. That is, an application 
of the analytic continuation formula 
of the hypergeometric function in terms of $x$ or $y$ to (2.12) yields 
the expression in some region we want. In this case 
 we should pay a little bit attention to deal with 
 logarithmic functions because the results of the following 
manipulations generally  depend on the choice of the branch 
 of the logarithmic function. Let us see 
how the solutions are transformed to 
 other Appell functions in various regions by these analytic properties. 
 If we take the branch for large $v$ in the weak coupling region, 
by taking inversion of the 
 variable $27v^2/4u^3$ of the expression (2.7)-(2.10), 
 we can give the solutions in the form of $F_4$ function, to be
 precise, $F_4(-1/6,1/3;2/3,1;4u^3/27v^2,\La^6/v^2)$ and its 
 independent solutions. Furthermore this $F_4$ function 
 can be written by $H_4$ function \cite{Erde}
\bea
H_4(a,b,c,2b;z,w)=
(1-{1\over 2}w)^{-a}F_4({1\over 2}a,{1\over 2}a+{1\over 2}
,c,b+{1\over 2};{16z\over (2-w)^2},{w^2\over (2-w)^2}),
\ena
where $a=-1/3,\ b=1/2,\ c=2/3,\ z=(\pm v+\La^3)^2u^3/108,\ w=
2\La^3/(\pm v+\La^3)$, and $H_4$ is another 
Appell function defined as 
\bea
H_4(\a,\b,\c,\d;z,w)=\sum_{m,n}{(\a)_{2m+n}(\b)_{n}\over 
(\c)_m(\d)_n\,m!n!}z^mw^n.
\ena
Notice that this function also can be written by 
 the hypergeometric function with $z$ or $w$. 
Although the formula (2.14) looks like 
the transformation about both two variables, only one 
variable $w$ is in fact transformed by using the quadratic 
transformation of the hypergeometric function. Similarly 
it is possible to transform the solutions by 
using well-known identities of the hypergeometric function, 
and in some cases they can be represented by Appell functions as 
the result of the transformation \cite{HTF,Erde}. 

 The most interesting region on the physical point of view is around 
 ${\bf Z}_3$ point $u=0,\ v=\pm \La^3$ where 
 the theory looks like the superconformal field theory \cite{AD}. 
 This point lies on the boundary of the convergence 
region of the expression $F_4$ for large $v$, and 
 already have appeared in the variables 
of $H_4$. Therefore 
 the solutions around this point can be given by 
 analytic continuation for $\La^6/v^2\sim 1$ from 
 $F_4$, or by the inversion of $w=2\La^3/(\pm v+\La^3)$ of $H_4$. After some 
 arrangements for the results to be simple forms, the solutions 
can be written by Appell $H_7$ functions. For detailed 
calculations, see ref.\cite{MS2}.  
 In this case, these analytic expressions show following 
properties; 
the logarithmic part of the solutions 
disappear after this analytic continuation and 
 Higgs field $a_i$ and its dual $a_D^i$ are on the equal 
standing up to overall constant, and 
 their leading dependence of moduli parameter 
becomes of fractional power \cite{MS2}. 
These properties indicate that the theory realize superconformal invariance 
 on this point as a nontrivial IR fixed point \cite{AD}. 

Next let us review how these four solutions satisfy the Picard-Fuchs equations
of this theory. The Picard-Fuchs equations for $\Pi =\oint \la$
are given by ${\cal L}_i \Pi =0 \   (i=0,1)$
where ${\cal L}_i$ is written by
\bea
{\cal L}_0&=&-3\partial^2_u+u\partial^2_v,\\
{\cal L}_1&=&4u^2\partial^2_u-9(\Lambda^6-v^2)\partial^2_v
+12uv\partial_u\partial_v+3v\partial_v+1.
\ena
By changing variables directly as $x=\frac{27\La^6}{4u^3},
y=\frac{27v^2}{4u^3}$ and setting $\Pi=\sqrt u F$,
we can write the differential equation for $F$ in the form as
$\tilde{\cal L}_i F =0 \  (i=0,1)$
where $\tilde{\cal L}_i$ is given by 
\bea
\tilde{\cal L}_0
&=&-x^2\partial_x^2-2xy\partial_x\partial_y+y(1-y)\partial_y^2
-x\partial_x
+(\frac{1}{2}-y)\partial_y+\frac{1}{36},\\
\tilde{\cal L}_1
&=&x(1-x)\partial_x^2-2xy\partial_x\partial_y-y^2\partial_y^2
+(1-x)\partial_x-y\partial_y+\frac{1}{36}.
\ena
This is Appell $F_4$ system constructed from the definition \cite{HTF}, 
whose independent four solutions turn out to be 
just the solutions derived in this section by linear combinations. 


\sect{$B_2$ and $C_2$ gauge theory}

In this section we perform the evaluation in the case of 
$B_2(SO(5))$ gauge theory at first, and $C_2(Sp(4))$ in the last of 
this section whose 
results can be given by a reparameterization of the results of $B_2$ case. 
Hyperelliptic curve and meromorphic differential $\la$ of $B_2$ theory 
are given by 
\bea
y^2&=&W(x)^2-\La^6 x^2,\ \ \ \ W(x)=x^4-ux^2+v,\label{eq:B2cur}
\ena
\bea
\la&=&{xW'(x)-W(x)\over y}dx\nonumber \\
 &=&dx \int_{-i\infty}^{i\infty}{ds\over 
2\pi i}{\G({1\over 2}+s)\G(-s)(-1)^s(\La^6)^6\over \G({1\over 2})\ 2s
}\left({x^2\over W(x)^2}\right)^s,
\ena
where we expand $\la$ in the weak coupling region,
  and introduce $s$ integral, and integrate by parts. In this case, 
we set $\beta_i$ cycle to intersect $\alpha_i$ cycle and 
its copy $\alpha_i'$ cycle both, 
because curve (3.1) is even function in terms of 
$x$. In the region $u
 \sim \infty$, $a_1$ consists of the contribution from 
$x^2\sim v/u$, and $a_2$ consists of from $x^2\sim u$. Therefore $a_1$ can be 
obtained by expanding $\la$ with respect to $1/(ux^2-v)$ and performing 
 $x$ integral of $\la$ 
from $x=-\sqrt{v/u}$ to $x=\sqrt {v/u}$ divide by 2, and 
 multiplied by $\sin 2\pi s/\pi$ so as to pick up the simple pole to 
 evaluate integration along $\alpha_1$. Corresponding $a_{D}^1$ is obtained 
 by integrating without multiplied 
by $\sin{2\pi s}$ in place of integration along $\beta_1$, 
 by evaluating double poles and by subtracting 
excessive contributions. The result can be written by
\bea
a_1&=&\sqrt{v\over u}\sum_{m,n}
{({1\over 2})_{n+2m}
(-{1\over 2})_{n-m}\over n!m!(1)_n}\left({v\over u^2}\right)^m
\left(-{\La^6\over 4uv}\right)^n\nonumber \\
&=&\sqrt{v\over u}H_5\left({1\over 2},-{1\over 2},1;
{v\over u^2},-{\La^6\over 4uv}\right),\\
a_D^1&=&{1\over 2\pi i}\sqrt{v\over u}\sum_{m,n}{({1\over 2})_{n+2m}
(-{1\over 2})_{n-m}\over n!m!(1)_n}\left({v\over u^2}\right)^m
\left(-{\La^6\over 4uv}\right)^n\nonumber \\
& & \times\left[
-2\psi(n+1)+\psi(n+2m+{1\over 2})+\psi(-{1\over 2}+n-m)+
\log\left(-{\La^6\over 4uv}\right)\right].
\ena
These are Appell $H_5$ functions which are order two hypergeometric 
series in terms of each variables defined by \cite{HTF}
\bea
H_5(a,b,c;x,y)=\sum_{m,n}{(a)_{n+2m}(b)_{n-m}\over (c)_nm!n!}
x^my^n,
\ena
which can be represented by using hypergeometric functions 
in terms of $-x$ and $y$ respectively. 
Contrast to $F_4$ function in 
the case of $SU(3)$, other solutions are not necessary represented 
by order two hypergeometric series as we will see in the following. 
By similar evaluations, we can get the expression for $a_2$ and $a_D^2$ by 
expanding $\la$ with $1/(ux^2-v)$ and by 
replacing lower and upper value of the 
 line integral to $x=-\sqrt u$ and $x=\sqrt u$ respectively 
\bea
a_2&=&u^{1\over 2}\sum_{m,n}{(-{1\over 2})_{3m+2n}\over (
{1\over 2})_{m+n}m!m!n!}\left({\La^6\over 4u^3}\right)^m\left(
-{v\over u^2}\right)^n, \\
a_D^2&=&{u^{1\over 2}\over 2\pi i}\sum_{m,n}{(-{1\over 2})_{3m+2n}\over
({1\over 2})_{m+n}m!m!n!}\left({\La^6\over 4u^3}\right)^m\left(
-{v\over u^2}\right)^n\nonumber \\
 & &\times\left[2\psi(m+1)-\psi({1\over 2}+m+n)+3\psi(-{1\over 2}+3m+2n)
+\log\left(-{\La^6\over 4u^3}\right)\right].
\ena
Thus we see that these solutions are of order three in terms of 
one variable. Although analytic property of the expression (3.6) and  
(3.7) are less manifest, we can see the behavior in 
various regions by using analytic property of the $H_5$ function 
(3.4) and (3.5). 

Next let us see how these four solutions can satisfy the Picard-Fuchs 
equations of this theory. Picard-Fuchs equations for $\Pi=\oint\la$ are 
given by ${\cal L}_i \Pi=0\ (i=0,1)$ where ${\cal L}_i$ is written by
\bea
{\cal L}_0&=&3\pa_u^2+u\pa_u\pa_v-v\pa_v^2,\\
{\cal L}_1&=&4u^2\pa_u^2-(9\La^6-16uv)\pa_u\pa_v+16v^2\pa_v^2+8v\pa_v+1.
\ena
By changing variables directly as $x={v\over u^2},\ y=-{\La^6\over 4uv}$ and 
setting $\Pi=\sqrt{v\over u}H$, we can write the differential equations 
for $H$ in the form as $\tilde{\cal L}_i H=0\ (i=0,1)$ where 
$\tilde{\cal L}_i$ are given by
\bea
\tilde{\cal L}_0&=&x(1+4x)\pa_x^2-y(1-4x)\pa_x\pa_y+y^2\pa_y^2+
({1\over 2}+8x)\pa_x+3y\pa_y+{3\over 4},\\
\tilde{\cal L}_1&=&2x^2\pa_x^2-xy\pa_x\pa_y+y(1-y)\pa_y^2+
{7\over 2}x\pa_x+(1-y)\pa_y+{1\over 4}.
\ena
These are the differential equations of 
Appell function $H_5$ constructed from the definition \cite{HTF}, 
whose independent four solutions are 
just our solutions obtained in this section. 

In the rest of this section we consider the case with gauge 
group $C_2$. It is known that correct holomorphic one-form and meromorphic 
differential can be obtained from 
spectral curve of the integral system $C_2^{\vee}$ defined by \cite{MW}
\bea
f=\left(z-{\mu\over z}\right)^2+x^2W(x)=0,\ \ \ W(x)=x^4-ux^2+v.
\ena
Meromorphic differential 
$\la$ which we use to get dual Higgs pairs in the gauge theory is given 
 from this curve by 
\bea
\la=x{dz\over z}.
\ena
Regarding $\pa_v \la$ as $dx/y$, this manipulation corresponds 
to take hyperelliptic curve as
\bea
y^2=x^4W(x)^2-\La^6x^2W(x),
\ena
where we set $\mu\propto \La^6$. Compared to curve (\ref{eq:B2cur})
this curve does not seem to
 reflect the property of group theory $SO(5)\sim Sp(4)$. 
 As was pointed out by Ito and Sasakura \cite{IS}, 
the result of $C_2$ is obtained from results of $B_2$ 
 by shifting moduli parameters as 
\bea
u^{C}=u^{B},\ \ v^{C}=v^{B}+{1\over 4}(u^B)^2.
\ena
This is observed by noticing 
that in this parameterization the Picard-Fuchs equations 
of both theories coincide \cite{IS}. Thus the solution of $C_2$ theory are 
given by Appell $H_5$ system with variables constructed by $u^{C},v^{C}$. 
 This is also observed by explicit evaluation in a same way as in the 
case of $B_2$ by 
using curve (3.14), and analytic continuation with respect to this 
change of parameterization. For example $a_1$ can be written as
\bea
a_1&=&\sum_{m,n}\sqrt{v^{C}\over u^{C}}{({1\over 2})_m({1\over 2})_{2n-m}\over 
({3\over 2})_{n-2m}m!m!n!}\left({-\La^6_Cu^{C}\over v^{C2}}\right)^m
\left({v^{C}\over u^{C2}}\right)^n\nonumber \\
&=&\sqrt{v^{C}\over u^{C}}\sum_{m}
{(-1)^m\over ({3\over 2})_{-2m}m!m!}\left(-{\La_C^6u^{C}
\over v^{C2}}\right)^mF(-{m\over 2}+{1\over 4},
-{m\over 2}+{3\over 4},-2m+{3\over 2};{4v^{C}\over u^{C2}}).
\ena
This function is not the Appell function, and similarly 
other three independent solutions are not within Appell functions. 
However by analytic continuation of the hypergeometric function
\bea
{1\over \G(c)}F(a,b,c,z)
&=&{\G(a+b-c)\over \G(a)\G(b)}(1-z)^{c-a-b}F(c-a,c-b,c-a-b+1;
1-z)\nonumber \\
 & & +{\G(c-a-b)\over \G(c-a)\G(c-b)}F(a,b,a+b-c+1;1-z),
\ena
and by use of the identity
\bea
F(a,b,c;z)=(1-z)^{c-a-b}F(c-a,c-b,c;z),
\ena
it is possible to get the expression in terms $u^{B}$ and $v^B$ as
a linear combination of the solutions of $B_2$
\bea
a_1&=&{1\over 2\sqrt{2\pi}}
\sqrt{v^{B}\over u^B}H_5\left({1\over 2},-{1\over 2},1;
{v^B\over u^{B2}},-{\La^6_B\over 4u^Bv^B}\right)\nonumber \\
 & &\hspace{2cm}+{1\over \sqrt{2\pi}}
\sqrt{u^{B}}\sum_{m,n}{(-{1\over 2})_{3m+2n}\over (
{1\over 2})_{m+n}m!m!n!}\left({\La^6_B\over 4u^{B3}}\right)^m\left(
-{v^{B}\over u^{B2}}\right)^n,
\ena
where we set $\La_B^6=-8\La_C^6$. Similarly we can evaluate 
to give $a_2,\ a_D^1$ and $a_D^2$ as some linear combinations 
of the result of $B_2$. 
Alternatively we can 
also verify by evaluations in a similar way initially by using this 
shifted parameterization as
\bea
\widetilde{W}(x) = x^4-u^{B}x^2+{1\over 4}u^{B2}+v^{B}.
\ena


\sect{$SO(4)$ theory}

In this section we deal with the $SO(4)$ theory whose hyperelliptic curve 
is given by
\bea
y^2=W(x)^2-\La^4x^4,\ \ W(x)=x^4-ux^2+v.
\ena
In relating to the change of the curve from the $B_2$ case, meromorphic 
differential $\la$ is varied slightly as 
\bea
\la&=&{xW'(x)-2W(x)\over y}dx\nonumber \\
&=&dx\int_{-i\infty}^{i\infty}{ds\over 2\pi i}{\G({1\over 2}+s)
\G(-s)(-\La^4)^s\over \G({1\over 2})2s}\left({x^4\over W(x)^2}
\right)^s.
\ena
Since the modification from the $B_2$ theory is only powers of 
$x$ in the instanton term, 
we are able to evaluate the integral in the region $\La\sim 0,\ u\sim \infty$
 in a same way as in the $B_2$ case. The result can be 
expressed by 
\bea
a_1&=&\sqrt{u}F_4\left(-{1\over 4},{1\over 4},{1\over 2},1;
{4v\over u^2},{\La^4\over u^2}\right), \\
a_2&=&\sqrt{v\over u}F_4\left({1\over 4},{3\over 4},
{3\over 2},1;{4v\over u^2},{\La^4\over u^2}\right),\\
a_D^1&=&{\sqrt u\over 2\pi i}\sum_{m,n}{(-({1\over 4})_{m+n}
({1\over 4})_{m+n}\over ({1\over 2})_m(1)_nm!n!}\left({4v\over u^2}\right)^m
\left({\La^4\over u^2}\right)^n\nonumber \\
 & & \times\left[
-2\psi(n+1)+\psi(n+m-{1\over 4})+\psi(n+m+{1\over 4})+\log\left(
{\La^4\over u^2}\right)\right],\\
a_D^2&=&{1\over 2\pi i}\sqrt{v\over u}\sum_{m,n}{({1\over 4})_{m+n}
({3\over 4})_{m+n}\over ({1\over 2})_m(1)_nm!n!}\left({4v\over u^2}\right)^m
\left({\La^4\over u^2}\right)\nonumber \\
& &\times \left[-2\psi(n+1)+\psi({1\over 4}+m+n)+\psi({3\over 4}+m+n)+
\log\left({\La^4\over u^2}\right)\right].
\ena
Again these solutions are written by Appell $F_4$ functions. 
 It is interesting to see how this result reflect the fact $SO(4)\sim 
SU(2)\times SU(2)$. To this end, we use 
following identities to relate $F_4$ function 
to the usual hypergeometric function 
\bea
F_4\left(\alpha,\alpha+{1\over 2},{1\over 2},\gamma;x,y\right)&=&
{(1+\sqrt x)^{-2\alpha}\over 2}F\left(\alpha,\alpha+{1\over 2},
\gamma;{y\over (1+\sqrt x)^2}\right)\nonumber \\
& &+{(1-\sqrt x)^{-2\alpha}\over 2}F\left(\alpha,\alpha+{1\over 2},\gamma;
{y\over (1-\sqrt x)^2}\right),\\
F_4\left(\alpha+{1\over 2},\alpha+1,{3\over 2},\gamma;x,y\right)&=&
{-x^{-{1\over 2}}\over 4\alpha}\left\{
(1+\sqrt x)^{-2\alpha}F\left(\alpha,\alpha+{1\over 2},
\gamma;{y\over (1+\sqrt x)^2}\right)\right.\nonumber \\
& &-\left.(1-\sqrt x)^{-2\alpha}F\left(\alpha,\alpha+{1\over 2},\gamma;
{y\over (1-\sqrt x)^2}\right)\right\}.
\ena
The first formula is cited in Ref. \cite{AK}, 
and second formula has been obtained in Ref. \cite{Su}. Applying 
these identities with $\alpha=-1/4$, $a_1$ can be expressed by using two 
same hypergeometric functions  
 of different variables as 
\bea
a_1&=&{1\over 2}(u+\sqrt{4v})^{1\over 2}F\left(
{1\over 4},-{1\over 4},1;{\La^4\over (u+\sqrt{4v})^2}\right)
\nonumber \\
 & & +{1\over 2}(u-\sqrt{4v})^{1\over 2}F\left(
{1\over 4},-{1\over 4},1;{\La^4\over (u-\sqrt{4v})^2}\right),
\ena
where $a_2$ can be given by changing sign of second term. 
Noticing that in the case of $SU(2)$ theory Higgs field $a$ is written by 
\cite{SW}
\bea
a=\sqrt uF\left({1\over 4},-{1\over 4},1;{\La^4\over u^2}\right),
\ena
it is recognized that the solution of $SO(4)$ theory decomposes to 
 two solutions of independent $SU(2)$ theories. This manifestly 
respects the fact $SO(4)\sim SU(2)\times SU(2)$. 

We can also rewrite the Picard-Fuchs equations of the theory ${\cal L}_i
\Pi=0\ 
(i=0,1)$ to Appell $F_4$ system where $\Pi=\oint \la$, and 
${\cal L}_i$ are given by 
\bea
{\cal L}_0&=&\pa_v+2\pa_u^2+2v\pa_v^2,\\
{\cal L}_1&=&-4(\La^4-u^2)\pa_u^2+16uv\pa_u\pa_v+16v^2\pa_v^2+8v\pa_v+1.
\ena
Changing variables as $x=-4v/u^2,\ y=\La^4/u^2$ and 
taking some linear combinations, we can see that the Picard-Fuchs 
equations reduce to Appell $F_4$ system $\tilde{\cal L}_i H=0$ 
where $H=u^{-{1\over 2}}\Pi$ and $\tilde{\cal L}_i$ are given by
\bea
\tilde{\cal L}_0=x(1-x)\pa_x^2-2xy\pa_x\pa_y-y^2\pa_y^2+({1\over 2}-x)\pa_x
-y\pa_y+{1\over 16},\\
\tilde{\cal L}_1=-x^2\pa_x^2-2xy\pa_x\pa_y+y(1-y)\pa_y^2-
x\pa_x+(1-y)\pa_y+{1\over 16}.
\ena
of the Picard-Fuchs equations. 
Let us see how the group property $SO(4)\sim SU(2)\times 
SU(2)$ can be realized from the point of view of the Picard-Fuchs equations. 
 In this case, we choose the variables as 
 $z=\La^4/(u+\sqrt{4v})^2$ and $w=\La^4/(u-\sqrt{4v})^2$ 
 so that the Picard-Fuchs equations defined by (4.11) and (4.12) can be 
combined to following two differential equations. 
One is $\pa_z\pa_w\Pi=0$, which means that 
$\Pi$ can be decomposed to the function $f(z)$ and 
$g(w)$ separately. By setting $f_0(z)=z^{1/4}f(z)$ and $g_0(w)=w^{1/4}
g(w)$, the second equation can be written as
\bea
& &z^{-{1\over 4}}
\left( (1-z)z\pa_z^2+(1-z)\pa_z+{1\over 16}\right)f_0(z)\nonumber \\
& & \hspace{3cm} +w^{-{1\over 4}}
\left((1-w)w\pa_w^2+(1-w)\pa_w+{1\over 16}\right)g_0(w)
=0.
\ena
Notice that 
famous Picard-Fuchs equation for $\Pi=\oint \la$ in
 the case of $SU(2)$ without matters \cite{SW}, 
 can be transformed by using the variable $z=\La^4/u^2$ 
 and setting $\Pi=u^{1/2}\Pi_0$ as follows
\bea
\left(4(u^2-\La^4)\pa_u^2+1\right)\Pi=0
\longrightarrow 
\left( (1-z)z\pa_z^2+(1-z)\pa_z+{1\over 16}\right)\Pi_0=0, 
\ena
which is just the hypergeometric differential equation. 
Differential equation (4.15) and (4.16) manifestly show that 
 the result of $SO(4)$ should be represented by some combinations of 
 the solutions of two independent $SU(2)$ theories.

\sect{Integrable models of type $C_2$}

In this section, we deal with the theory with respect to 
the integrable system $C_2$. The spectral curve of this system is given by 
\cite{MW}
\bea
f=z+{\mu\over z}+W(x)=0,\ \ \ W(x)=x^4-ux^2+v.
\ena
Setting the variable $y=z+W(x)/2$ and $\mu\propto \La^8$, 
we can get the hyperelliptic 
curve of corresponding Seiberg-Witten model in the following form:
\bea
y^2=W(x)^2-\La^8.
\ena
However, looking at the simple singularity part $W(x)$ which corresponds to 
classical singularity of gauge theory, and counting 
the dimension of $\mu$ which relate to the power of $\La$ 
of instanton corrections, the theory constructed from 
this curve does not correspond to any known gauge theory. This theory 
does not seem to have any lagrangian description, in other 
words, it may describe the purely strong coupling theory. 
Since our interest 
is the connection between Seiberg-Witten theories with rank two and 
hypergeometric series, it is worth seeing how this theory 
can be represented by hypergeometric series. 
Meromorphic differential $\la$ is given in the weak coupling region by
\bea
\la&=&xW'(x){dx\over y}={xW'(x)dx\over W(x)}\sum_{n=0}^{\infty}
{\G(n+{1\over 2})\over \G({1\over 2})n!}\left({
\La^8\over W(x)^2}\right)^n \nonumber \cr
&=& dx\int {ds\over 2\pi i}{\G(s+{1\over 2})\G(-s)(-1)^s\over \G({1\over 2})
\  2s}\left({\La^8\over W(x)^2}\right)^s.
\ena
Since $W(x)$ is not varied from the $B_2$ and $C_2$ theory, we 
can evaluate in a same way as before. 
The result can be written as
\bea
a_1&=&u^{1\over 2}\sum_{m,n}{\G({1\over 2})\G(4n+2m-{1\over 2})
\over \G(-{1\over 2})\G(2n+m+{1\over 2})n!n!m!}
\left({\La^8\over 4u^4}\right)^n \left(-{4v\over u^2}\right)^m,\\
a_2&=&\sqrt{v\over u}\sum_{m,n}{\G(2n-m-{1\over 2})\G(m+{1\over 4})
\G(m+{3\over 4})\over \G(-{1\over2})\G({1\over 4})\G({3\over 4})n!n!m!}
\left({\La^8\over 4v^2}\right)^n\left(-{4v\over u^2}\right)^m 
\nonumber \\
 &=&\sqrt{v\over u}H_7\left(-{1\over 2},{1\over 4},{3\over 4},1;
{\La^8\over 4v^2};-{4v\over u^2}\right),\\
a_D^1&=&{u^{1\over 2}\over 2\pi i}\sum_{m,n}{\G({1\over 2})\G(4n+2m-{1\over 2})
\over \G(-{1\over 2})\G(2n+m+{1\over 2})n!n!m!}
\left({\La^8\over 4u^4}\right)^n \left(-{4v\over u^2}\right)^m\nonumber \\
 & &\times\left[-2\psi(n+1)+4\psi(4n+2m-{1\over 2})-
2\psi(2n+m+{1\over 2})+\log\left({\La^8\over 4u^4}\right)\right],\\
a_D^2&=&{1\over 2\pi i}
\sqrt{v\over u}\sum_{m,n}{\G(2n-m-{1\over 2})\G(m+{1\over 4})
\G(m+{3\over 4})\over \G(-{1\over2})\G({1\over 4})\G({3\over 4})n!n!m!}
\left({\La^8\over 4v^2}\right)^n\left(-{4v\over u^2}\right)^m \nonumber \\
 & &\times\left[-2\psi(n+1)+2\psi(2n-m-{1\over 2})+\log
\left({\La^8\over 4v^2}\right)\right],
\ena
where the Appell function $H_7$ is defined as \cite{HTF}
\bea
H_7(a,b,c,d;x;y)=\sum_{m,n}{(a)_{2n-m}(b)_m(c)_m\over 
(d)_nn!m!}x^ny^m.
\ena
Certainly it is easy 
to see the analytic property 
of this function. 

Let us see how these solutions can satisfy  the Picard-Fuchs equation.
 Picard-Fuchs equation for $\Pi=\oint \la$ is given by 
\bea
{\cal L}_0\Pi&=&\left(4\pa^2_u+2u\pa_u\pa_v+\pa_v\right)\Pi=0,\\
{\cal L}_2\Pi&=&\left((-4u^2+32v)\pa^2_u+16(\La^8-v^2)\pa_v^2-1\right)\Pi=0.
\ena
By direct change of the variable to $x=\La^8/ 4v^2$ and 
$y=-4v/u^2$, and some linear combinations, 
we see that these equations become Appell system $H_7$
 for $\Pi_0=\sqrt{u/v}\ \Pi$ as follows
\bea
& &\left (
y(1+y)\pa_y^2-2x\pa_x\pa_y+({3\over 2}+2y)\pa_y+{3\over 16}
\right)\Pi_0=0,\\
& &\left(-(4x-1)x\pa_x^2-y^2\pa_y^2+4xy\pa_x\pa_y+
(-4x+1)\pa_x-y\pa_y+{1\over 4}\right)\Pi_0=0.
\ena
Thus we see that 
$\Pi_0$ is just a Appell function $H_7$ and dual Higgs pairs can be 
represented by independent solutions of this system, which are just
the solution we have obtained in this section. 

It is also possible to consider other parameterization of 
the curve as
\bea
\tilde{u}=u,\ \ \ \tilde{v}=-v+{u^2\over 8}.\label{eq:intC2}
\ena
In this case, solutions are obtained from previous solutions by 
changing variables and analytic continuations. 
However since the result of this manipulation 
 depend on the choice of the 
branch of the logarithmic function, we 
instead take a different parameterization of the 
curve initially and perform the evaluation. $W(x)$ now can be written as 
\bea
\widetilde{W}(x)=\left(x^2-{1\over 2}(1+{1\over \sqrt 2})\tilde{u}\right)
\left(x^2-{1\over 2}(1-{1\over \sqrt 2})\tilde{u}\right)-\tilde{v},
\ena
therefore we regard $a_1$ as the contribution from $x^2\sim {1\over 2}(
1+{1\over \sqrt{2}})\tilde{u}
$ and $a_2$ from $x^2\sim {1\over 2}(1-{1\over \sqrt 2})\tilde{u}$. 
Expanding $\la$ suitably and integrating from $x=0$ to $x=\sqrt{
{1\over 2}(1\pm{1\over \sqrt 2})\tilde{u}}$ with appropriate 
regularization, we can give the result by using $F_4$ function. 
For example $a_1$ can be written as
\bea
a_1&=&{\sqrt{(1+{1\over \sqrt{2}})\tilde{u}}\over \sqrt 2}
F_4\left(-{1\over 8},{1\over 8},1,{1\over 2};
{64\La^8\over \tilde{u}^4},{64\tilde{v}^2\over \tilde{u}^4}\right)\\
& &\hspace{3cm}+
{\sqrt{2-\sqrt 2}
\tilde{v}\over \sqrt {2}\tilde{u}^{3\over 2}}F_4\left({3\over 8},{5\over 8},1,
{3\over 2};{64\La^8\over \tilde{u}^4},{64\tilde{v}^2\over \tilde{u}^4}
\right).\nonumber 
\ena
It is possible to obtain $a_2$ and $a_D^1, \ a_D^2$ as linear 
combinations of $F_4$ functions similarly. We can also
check these solutions can satisfy the Picard-Fuchs equation directly 
by choosing variables as in (\ref{eq:intC2}).

There is another way to express dual Higgs pairs by 
 Appell functions. 
If we consider the branch for large $v$ in the original parameterization, 
the result can be 
expressed in terms of $H_4$ function \cite{HTF} by similar evaluations. 
This situation in which dual Higgs pairs can be 
 represented by several kinds of Appell functions in each region is analogous 
to the relation between the weak coupling region and around the 
conformal point in the case of $SU(3)$ \cite{MS2}. 

\sect{Exceptional group $G_2$}

In this section we deal with the theory with gauge group $G_2$. As 
was pointed out in \cite{AAMLPG,AAG},
 since there is no reliable hyperelliptic curve for $G_2$ so far,
  we start with spectral curve for integrable systems. 
The spectral curve for $G_2^{\vee}$ is given by \cite{MW}
\bea
f&=&3\left(z-{\mu\over z}\right)-x^8+2ux^6-\left[u^2+6\left(z+{\mu\over z}
\right)\right]x^4+\left[v+2u\left(z+{\mu\over z}\right)\right]x^2 
\nonumber \\
&=&3\left(z-{\mu\over z}\right)^2+\left(z+{\mu\over z}\right)
\left({u\over 3}-x^2\right)6x^2-P(x).\label{eq:G2holo}
\ena
where
\bea
P(x)=x^2\left[x^2(x^2-u)^2-v\right].
\ena
Holomorphic one-form for periods is given from this spectral curve by
\bea
\int {dxdz\over z\cdot f}.\label{eq:G2curv}
\ena
Integrating with respect to $v$ we can get
meromorphic differential which produces dual Higgs pairs of corresponding
Seiberg-Witten model.
Expanding with respect to $1/P(x)$, holomorphic one-form is given by
\bea
\int{dxdz\over zf}=dx\sum_{n,k}{\G(n+k+1)\ x^{4k}({u\over 3}-x^2)^{2k}
(4\mu)^n\over \G(n-k+1)\G(k-n+{1\over 2})n!k!}\left({P(x)\over 3}
\right)^{-(n+k+1)}.\label{eq:G2holo2}
\ena
The detailed procedure is given in Appendix B. 
Thus meromorphic differential $\la$ can be written as follows
\bea
\la=dx\int{ds\over 2\pi i}\sum_{k}{\G(s+k)\G(-s)(-1)^s(4\mu)^sx^{4k}
({u\over 3}-x^2)^{2k}\over \G(s-k+1)\G(k-s+{1\over 2})k!}
\left({P(x)\over 3}\right)^{-(s+k)}.
\ena
In the region $u\sim \infty$, we expand this expression with 
respect to $1/u$ and deform the contour to evaluate $x$ integral 
for Higgs pairs explicitly. By picking up poles $s=0,1,\cdots $,
 the contribution from $x^2\sim u$ and 
$x^2\sim u/u^2$ to 
lower orders of $\mu$, which amount to $a_1$ and $a_2$ respectively, turn 
out to be completely match the result obtained by Ito \cite{I}. 
However it is difficult to 
combine this expression to simple known hypergeometric functions. Therefore
 we choose a following parameterization $\tilde{u}, 
\tilde{v}$ which makes the calculation simple 
enough to evaluate analytically
\bea
\tilde{u}=u,\ \ \ v={4\over 27}u^3+\tilde{v}.
\ena
In this parameterization, 
 $P(x)$ shows the classical 
singularity as
\bea
P(x)=x^2\left[\left(x^2-{\tilde{u}\over 3}\right)^2\left(x^2-{4\tilde{u}
\over 3}\right)-\tilde{v}\right].
\ena
We are going to evaluate the contribution from $x^2\sim \tilde{u}/3$ and 
$x^2\sim 4\tilde{u}/3$. At first we attempt to calculate 
the contribution from $x^2\sim 4\tilde{u}/3$, however  
it is still difficult to combine the result to simple form. Then 
we instead try to calculate $a_1$ from the contribution from 
$x^2\sim \tilde{u}/3$. Since this point is a double root of $P(x)=0$, 
the result of the calculation consist of one independent solution only, 
which we call even part $a^{+}$. 
Expanding meromorphic differential (6.5) in terms 
of large $\tilde{u}$, performing the line integral from 
$x=-\sqrt{\tilde{u}/3}$ to $x=\sqrt{\tilde{u}/3}$ multiplied 
by $\sin 2\pi s/\pi$, we can give 
the solution in the following form:
\bea
a_1=a^{+}=u^{1\over 2}\sum_{m,n}{\G({1\over 2})\G(4m+3n-{1\over 2})\over 
\G(-{1\over 2})\G(m+n+{1\over 2})^2m!m!n!}\left({9\mu\over 
\tilde{u}^4}\right)^m\left({\tilde{v}\over 4\tilde{u}^3}\right)^n
.\label{eq:G2a1}
\ena
The derivation of this expression is explained
 in detail in Appendix B. Another polynomial solution which we call 
 odd part $a^{-}$ can be given by shifting $n$ to $n-m+1/2$ as 
\bea
a^{-}={\tilde{v}^{1\over 2}\over \tilde{u}}\sum_{m,n}{\G({3\over 2})
\G(m+3n+1)\over \G(n-m+{3\over 2})m!m!n!n!}\left({36\mu\over \tilde{u}
\tilde{v}}\right)^m\left({\tilde{v}\over 4\tilde{u}^3}\right)^n,
\label{eq:G2a2}
\ena
which is because of the fact that this replacement does not 
change the recursion relation satisfied by the coefficients. The solution 
 (\ref{eq:G2a1}) and 
(\ref{eq:G2a2}) both have forms of hypergeometric series with 
order three in terms of 
one variable $\tilde{v}/4\tilde{u}^3$. Thus we see that 
in the case of $G_2$ dual Higgs pairs are not within the Appell system. 
If we can evaluate the contribution from $x^2\sim 4\sqrt{\tilde{u}/3}$, 
we would see that $a_2$ can be written by a linear combination of 
$a^{+}$ and $a^{-}$. Similarly we can give the logarithmic solutions $a_D^{+}$ 
and $a_D^{-}$. An order three hypergeometric function satisfies
 suitable order three differential equation by definition. 
Odd part solution $a^{-}$ 
 is of order 2 in terms of $36\mu/\tilde{u}\tilde{v}$, and 
of order 3 in terms of $\tilde{v}/4\tilde{u}^3$, therefore 
two more independent solutions seem to be needed. 
However looking at the form of these solutions, we recognize that 
the number of independent solution of this system is four including 
logarithmic solutions. This is implied by the fact that 
the Picard-Fuchs equations of this theory are order two differential 
equations \cite{I}. 

In the case of $G_2$, we have succeeded in evaluating 
dual Higgs pairs explicitly by means of parameterization (6.6) only, 
and have not been able to sum up to the simple form by use 
of other parameterizations so far. Thus 
we recognize that $G_2$ seems to be {\it exceptional} case 
compared to the classical gauge group where we can 
evaluate in various parameterizations. 

Let us check whether these solutions can satisfy 
 the Picard-Fuchs equations. Picard-Fuchs equations for $\Pi=\oint \la$ are
 given by ${\cal L}_i
\Pi=0\ (i=1,2)$ where \cite{I}
\bea
{\cal L}_1&=&\left({8u^3v\over 3}-36v^2+960u^2\mu\right)\pa^2_v+
\left({8u^4\over 3}-24uv+2304\mu\right)\pa_u\pa_v+(4u^3-24v)\pa_v-1,
\nonumber \\
{\cal L}_2&=&{2(720u^2\mu+2u^3v-27v^2)\over -uv+24\mu}\pa^2_u
+{4(256u^4\mu-3u^2v^2-720uv\mu+13824\mu^2)\over -uv+24\mu}\pa_u\pa_v
\nonumber \\
& &\hspace{2cm}-{6(-256u^3\mu+96v\mu+5uv^2)\over -uv+24\mu}\pa_v-1.
\ena
By direct change of the variable $x=36\mu/\tilde{u}\tilde{v},
\ y=\tilde{v}/4\tilde{u}^3$,
and applying (\ref{eq:G2a2}) to this system, 
we can check directly 
that the solution $a^{-}$ satisfies these Picard-Fuchs equations order 
by order. Also we can show that the recursion relations 
 with respect to these equations can be satisfied by  coefficient $c(m,n)$ in 
arbitrary integer $m,n$ where we set $a^{-}=\tilde{v}^{1\over 2}/
\tilde{u}\sum_{m,n}c(m,n)x^my^n$. Therefore 
we regard $a^{+},a^{-}$ and their corresponding logarithmic solutions as 
 four independent solutions of the Picard-Fuchs equations. 
However this occurs nontrivially in a sense that 
 we could not succeed to combine ${\cal L}_i$ to 
differential equations $\tilde{\cal L}_i
\Pi_0=0\ (i=1,2)$ where
\bea
\tilde{\cal L}_1&=&
(1+x)x\pa_x^2+2xy\pa_x\pa_y-3y^2\pa^2_y+(1+{3x\over 2})\pa_x-{11\over 2}y
\pa_y-{1\over 2},\nonumber \\
\tilde{\cal L}_2&=&-z^3\pa_x^3-(xy+27xy^2)\pa_x\pa_y^2-9x^2y\pa_x^2\pa_y+
(1-27y)y^2\pa_y^3-9x^2\pa_x^2 \\
& &+({7y\over 2}-162y^2)\pa_y^2-(x+72xy)\pa_x\pa_y-
18x\pa_x+({3\over 2}-114y)\pa_y-6,\nonumber 
\ena
which 
should be simply satisfied by $\Pi_0=\tilde{u}/\tilde{v}^{1\over 2}a^{-}$ by 
definition. This is analogous to the case of Calabi-Yau 
 \cite{HKTY} where 
 the system has the redundancy which appears as 
 the factorization of some differential operators. This feature 
 decreases the order of differential equations satisfied by the solution, 
 to give the essential Picard-Fuchs equations. 
  Unfortunately we have not succeed in finding suitable 
 parameterization which makes us possible to do 
 in practice so far. 

\sect{Conclusion}

We have showed how dual Higgs pairs for Seiberg-Witten 
theories with rank two gauge groups can be represented by 
hypergeometric series in the weak coupling region both 
by using explicit evaluations and by the Picard-Fuchs equations. 
In the case of classical gauge groups, 
they are written by Appell functions which is natural 
 extensions of the hypergeometric function in two variables. 
In the case of the exceptional group, however, we need to 
extend one more order, that is, they can be expressed by 
order three hypergeometric series in terms of one of two variables. 
Compared to classical groups this is realized nontrivially in a following 
sense that equations 
which should be satisfied by these solutions, look like independent 
from the Picard-Fuchs equations of the theory, although
 these solutions can satisfy the Picard-Fuchs equations. 
 
Explicit evaluations are not in principle 
suffered from the complexity coming from gauge group 
 being higher rank. However analytic property of each variable of 
the resulting expressions are not always apparent \cite{MS2}. 
 In order to clarify this we need more information 
about generalized hypergeometric functions with multi-variable. 
 We hope that this makes us possible to analyze 
 the behavior in the strong coupling region of 
 Seiberg-Witten type theories with arbitrary gauge groups, 
even within exceptional groups, in future.

\appendix

\section{Explicit evaluations in the case of classical group}

Consider the curve of following form:
\bea
y^2=W(x)^2-x^{k}\La^d.
\ena
Meromorphic differential is given by
\bea
\la={xW'(x)-{k/2}W(x)\over y}dx.
\ena
Dual Higgs pairs $a_i,\ a_D^i\ (i=1,2)$ are given by integration 
along independent suitable homology cycle $\alpha_i,\ \beta_i$ cycle 
respectively. In the weak coupling region $\La\sim 0$, 
we expand $\la$ with $1/W(x)^2$ 
\bea
\la=dx\left(xW'(x)-{k\over 2}W(x)\right)
\sum_{n}{\G(n+{1\over 2})\over \G({1\over 2})n!}
\left({\La^dx^k\over W(x)^2}\right)^n.
\ena
Introducing $s$ integral in place of summation in terms of $n$, and 
integrating by parts, we see that dual Higgs pairs consist 
of the contribution from classical root $W(x)=0$ of the curve as
\bea
\la=dx\int_{-i\infty}^{i\infty}
{ds\over 2 \pi i}{\G(s+{1\over 2})\G(-s)(-1)^s\over 
\G({1\over 2})\ 2s}\left({\La^dx^k\over W(x)^2}\right)^s,
\ena
where we take poles at 
$s=0,1,2,\cdots$. In the case of $C_2(Sp(4))$, we instead start 
with holomorphic one-form with respect to 
hyperelliptic curve $y^2=x^4W(x)^2-\La^6x^2W(x)$, or 
corresponding following form obtained from 
spectral curve $f=(z-{\mu\over z})^2+x^2W(x)=0$ as 
\bea
\int{dxdz\over z\cdot f}.
\ena
Analyzing the residue of $z$ integral we obtain same holomorphic one-form. 
 Integrating with respect to $v$ and expanding with $1/W(x)$, 
 we see that $\la$ can be written in a similar form as (A.4).

In the weak coupling region, since each root of hyperelliptic curve reduce 
to the classical root of $W(x)$, we can deform the contour for 
$\alpha_i$ and $\beta_i$ cycles appropriate way to enclose these roots. 
 $a_i$ and $a_D^i$ are obtained by evaluating the contribution 
from these points by picking up suitable poles with respect to 
corresponding cycles, so that this reduction gives correct 
asymptotic behaviors in the weak coupling region. 

 In the case of $SU(3)$ where $W(x)=x^3-ux-v$, in the region $u\sim \infty$, 
$\alpha_i\ (i=1,2,3)$ enclose point $x\sim \sqrt u,\ -\sqrt u,\ -v/u$ 
respectively. Contour integral along $\alpha_i$ cycle can be evaluated by 
using the line integral with suitable regularization to produce 
the polynomial solution. This is possible to multiply $\sin{2s\pi}/\pi$ 
to the expression and perform the line integral from one root to 
another root of $W(x)$ and evaluate simple pole of $s$ integral. 
For example, the solution which consist of the difference 
between the contribution from $x\sim \sqrt u$ and from 
$x\sim -v/u\sim 0$, which we call $a_1$, 
 is obtained by expanding $\la$ multiplied by 
$\sin{2s\pi}/\pi$ with large $u$ and by integrating from $x=0$ to 
$x=\sqrt u$ as \cite{MS2,MS}
\bea
a_1&=&\int_0^{\sqrt u}dx\int{ds\over 2\pi i}{\sin 2\pi s\over \pi}
\sum_{m}{\G(s+{1\over 2})\G(-s)(-1)^s\G(2s+m)\over
 \G({1\over 2})\G(2s+1)m!}(\La^6)^s
v^m\nonumber \\
 & &\hspace{2cm}\times x^{-2s-m}(x^2-u)^{-2s-m}
\ena
Performing line integral by changing variable as $x^2=ut$ and evaluating
 simple poles of  resulting expression, 
we can give the expression (\ref{eq:A2a1}) for $a_1$. 
To get corresponding $a_D^1$ which is obtained by integration along 
$\beta_1$, we integrate $\la$ similarly without 
multiplied by $\sin 2\pi s$ and 
 evaluate double poles and subtract $a_1/2$ \cite{MS2,MS}. 
The result can be written as (\ref{eq:A2aD1}). 
Replacing $\sqrt u$ to $-\sqrt u$, 
we can obtain  expression $a_2,\ a_D^2$ which contain 
the difference between the contribution from $x\sim -\sqrt u$ and from $x
\sim -v/u$.

 In $SO(5), SO(4)$ and the theory based on integrable model $C_2$ case, 
 we use $W(x)=x^4-ux^2+v$. The contribution comes from 
 $x^2\sim u$ and $x^2\sim v/u$. To evaluate the former contribution, 
we expand $\la$ with $1/(x^4-ux^2)$ and integrate in a similar way as 
$SU(3)$ case. To calculate latter contribution, we expand with 
 $1/(ux^2-v)$ and integrate from $x=0$ to $x=\sqrt{v/u}$, and 
 evaluate poles regulated
suitably as before. In this way we can derive the expression 
for $a_i, a_D^i$ of the theory with the classical gauge group. 

\sect{Explicit evaluations in the case of exceptional group}

In the case of $G_2$, the spectral curve of the integrable system $G_2^{\vee}$ 
is given by (\ref{eq:G2curv}). In the weak coupling region, 
we expand (\ref{eq:G2holo}) with respect to $1/P(x)$
\bea
\int{dxdz\over 2\pi i \,zf}&=&{dx\over 2\pi i}\int{dz\over z}\sum_{n}\left[
\left(z-{\mu\over z}\right)^2+2x^2\left(z+{\mu\over z}\right)\left({u\over 3}-
x^2\right)\right]^n\left({P(x)\over 3}\right)^{-n-1}\nonumber \\
&=&{dx\over 2\pi i}\int{dz\over z}\sum_{n,k}{n!\over k!(n-k)!}
\left(z-{\mu\over z}\right)^{2(n-k)}
\left(z+{\mu\over z}\right)^k \\
& &\hspace{4cm}\times\left[
2x^2\left({u\over 3}-x^2\right)\right]^k\left({P(x)\over 3}\right)^{-n-1}.
\nonumber 
\ena
Expanding both braces of integrant, we evaluate the residue at 
 $z=0$
\bea
& &{1\over 2\pi i}\int{dz\over z}\left(z-{\mu\over z}\right)^{2(n-k)}
\left(z+{\mu\over z}\right)^k \nonumber \\
&=&\sum_{i}{\left(2(n-2k)\right)!\ (2k)!\ \mu^{n-k}\ (-1)^i\over 
\G(2n-4k-i+1)\G(3k-n+i)\G(n-k-i+1)\ i!} \nonumber \\
&=&{(2k)!\ \mu^{n-k}\over\G(3k-n+1)\G(n-k+1)}F(4k-2n,k-n,3k-n+1;-1).
\ena
Using the value of the hypergeometric function \cite{HTF}
\bea
F(4k-2n,k-n,3k-n+1,-1)={2^{-4k+2n}\G(3k-n+1)\G({1\over 2})\over 
\G(k+1)\G(2k-n+{1\over 2})},
\ena
we can get expression (\ref{eq:G2holo2}) which 
corresponds to holomorphic one-form. Thus the meromorphic differential 
is given by 
\bea
\la=dx\int{ds\over 2\pi i}\sum_{k}{\G(s+k)\G(-s)(-1)^s(4\mu)^sx^{4k}
({u\over 3}-x^2)^{2k}\over \G(s-k+1)\G(k-s+{1\over 2})k!}
\left({P(x)\over 3}\right)^{-(s+k)},
\ena
where we introduce $s$ integral ($s=0,1,2,\cdots$) as before.

We take different parameterization convenient for 
evaluation of the integral as
\bea
P(x)=x^2\left[\left(x^2-{\tilde{u}\over 3}\right)^2\left(x^2-{4\tilde{u}
\over 3}\right)-\tilde{v}\right].
\ena
We are going to evaluate the contribution from $x^2\sim u/3$. To 
get the polynomial solution $a^{+}$, we calculate
\bea
a^{+}=\int_{-\sqrt{\tilde{u}\over 3}}^{\sqrt{\tilde{u}
\over 3}}\la{\sin 2\pi(s+k)\over 
\pi},
\ena
where we expand $\la$ with $1/(x^2-u/3)$. 
In order to perform $x$ integral we use integral representation for 
the hypergeometric function 
\bea
F(a,b,c;z)={\G(c)\over \G(b)\G(c-b)}
\int_0^1dt\ t^{b-1}(1-t)^{c-b-1}(1-zt)^{-a},
\ena
so that we give the following expression
\bea
a^{+}&=&\tilde{u}^{1\over 2}\int{ds\over 2\pi i}\sum_{k,m}
{\sin 2\pi(s+k)\over \pi}{\G(-s)(-1)^s\G(s+k+m)\over 
\G(s-k+1)\G(k+1)\G(m+1)}\nonumber \\
& &\hspace{1cm}\times {\G({3\over 2})\G(-2s-2m+1)\over \G(-3s+k-2m+{3\over 2})}
\left(4\mu\over \tilde{u}^4\right)^s\left(\tilde{v}\over 
\tilde{u}^3\right)^m{3^{5s+k+3m}\over (-4)^{s+k+m}}\\
& &\hspace{2cm}\times 
F(s+k+m,-s+k+{1\over 2},-3s+k-2m+{3\over 2};{1\over 4}).\nonumber 
\ena
We want to modify this expression to simple form. First of all 
 we evaluate the value of the hypergeometric function by using 
analytic continuation formula
\bea
F(a,b,c;z)&=&(1-z)^{-a}{\G(c)\G(b-a)\over \G(b)\G(c-a)}
F(a,c-b,a-b+1;{1\over 1-z})\\
& &\hspace{2cm}+(1-z)^{-b}{\G(c)\G(a-b)\over 
\G(a)\G(c-b)}F(b,c-a,b-a+1;{1\over 1-z}),\nonumber
\ena
where $a=s+k+m,\ b=-s+k+1/2,\ c=-3s+k-2m+3/2$ and $z=
1/4$. Notice that in this case there is no contribution 
from second term because poles with respect to $s$ disappear 
after this manipulation. Then we have
\bea
a^{+}&=&\tilde{u}^{1\over 2}\int{ds\over 2\pi i}
\sum_{k,m}{\sin 2\pi(s+k)\over \pi}
{\G(-s)(-1)^s\G(s+k+m)3^{4s+2m}(-1)^{s+k+m}
\over \G(s-k+1)\G(k+1)\G(k-s+{1\over 2})\G(m+1)}\nonumber \\
& &\hspace{1cm}\times 
{\G({3\over 2})\G(-2s-2m+1)\G(-2s-m+{1\over 2})\over \G(-4s-3m+{3\over 2})}
\left({4\mu\over \tilde{u}^4}\right)^s\left({\tilde{v}\over \tilde{u}^3}
\right)^m \\
& &\hspace{2cm}\times 
F(s+k+m,-2s-2m+1,2s+m+{1\over 2};{4\over 3}).
\ena
Next we intend to sum up with respect to $k$. Getting hypergeometric 
function back to the form of summation in term of $l$, 
and combining $k$ summation to hypergeometric function with unit value 
which is given by \cite{HTF}
\bea
F(a,b,c;1)={\G(c)\G(c-a-b)\over \G(c-a)\G(c-b)},
\ena
we give the following expression by rewriting again resulting 
 $l$ summation to hypergeometric function with value $4/3$
\bea
a^{+}&=&\tilde{u}^{1\over 2}\int{ds\over 2\pi i}\sum_{m}
{\G({1\over 2})\G(-s)(-1)^s\G(s+m)
\G(4s+3m-{1\over 2})3^{4s+2m}\over 
\G(-{1\over 2})\G(s+1)\G(m+1)\G(2s+2m)\G({1\over 2}+s+m)}
\nonumber \\
& & \hspace{2cm}\times \left({4\mu\over \tilde{u}^4}\right)^s
\left({\tilde{v}\over \tilde{u}^3}\right)^m
F(1-2s-2m,s+m;{1\over 2}+s+m;{4\over 3}).
\ena
Finally in order to evaluate the value of the hypergeometric function, 
we try to use next formula for arbitrary integer $n$ 
\bea
F(-2n+1,n;{1\over 2}+n;{4\over 3})={1\over 3^{2n}}.
\ena
This formula is not sited elsewhere, however we can verify directly 
by substituting small integer $n$, and for large $n$ by using mathematica. 
Using this formula, we can get $a^{+}$ in the form:
\bea
a^{+}=u^{1\over 2}\sum_{m,n}{\G({1\over 2})^2\G(4m+3n-{1\over 2})\over \G(-
{1\over 2})\G(m+n+{1\over 2})^2 m!m!n!}\left({9\mu\over u^4}\right)^m
\left({\tilde{u}\over 4u^3}\right)^n.
\ena

\newpage

\setlength{\baselineskip}{14pt}

\end{document}